\documentclass[%
 reprint,
superscriptaddress,
 amsmath,amssymb,
 aps,
 prl,
]{revtex4-1}

\usepackage{siunitx}
\usepackage{graphicx}% Include figure files
\usepackage{dcolumn}% Align table columns on decimal point
\usepackage{bm}% bold math
\usepackage{epstopdf}
\usepackage[hidelinks]{hyperref}

\makeatletter
\newcommand{\mypm}{\mathbin{\mathpalette\@mypm\relax}}
\newcommand{\@mypm}[2]{\ooalign{%
		\raisebox{.1\height}{$#1+$}\cr
		\smash{\raisebox{-.6\height}{$#1-$}}\cr}}
\makeatother

\makeatletter
\newcommand*{\citenumns}[2][]{%
	\begingroup
	\let\NAT@mbox=\mbox
	\let\@cite\NAT@citenum
	\let\NAT@space\NAT@spacechar
	\let\NAT@super@kern\relax
	\renewcommand\NAT@open{}%
	\renewcommand\NAT@close{}%
	\cite[#1]{#2}%
	\endgroup
}
\makeatother

\usepackage{commath}
\usepackage{tabularx}

\usepackage{booktabs}% for better rules in the table

\begin{document}

\preprint{APS/123-QED}

\title{Probing Electron Spin Resonance in Monolayer Graphene}% Force line breaks with \\

\author{T.J. Lyon}
\email{tlyon@wisc.edu}
\affiliation{%
	Center for Hybrid Nanostructures (CHyN), Department of Physics, University of Hamburg,
	Jungiusstrasse 9-11, 20355 Hamburg, Germany\\
}
\affiliation{Department of Physics, University of Wisconsin-Madison, 1150 University Ave, Madison, Wisconsin 53706, USA}%Lines break automatically or can be forced with \\
\author{J. Sichau}%
\author{A. Dorn}%
\affiliation{%
	Center for Hybrid Nanostructures (CHyN), Department of Physics, University of Hamburg,
	Jungiusstrasse 9-11, 20355 Hamburg, Germany\\
}%
\author{A. Centeno}
\author{A. Pesquera}
\author{A. Zurutuza}
\homepage{http://www.graphenea.com}
\affiliation{Graphenea, Avenida de Tolosa 76, 20018 Donostia-San Sebastian, Spain\\
}%
\author{R.H. Blick}%
\homepage{http://www.nanomachines.com}
\affiliation{%
	Center for Hybrid Nanostructures (CHyN), Department of Physics, University of Hamburg,
	Jungiusstrasse 9-11, 20355 Hamburg, Germany\\
}%

\date{\today}% It is always \today, today,

\begin{abstract}
The precise value of the $g$-factor in graphene is of fundamental interest for all spin-related properties and their application. We investigate monolayer graphene on a Si/SiO$_{2}$ substrate by resistively detected electron spin resonance (ESR). Surprisingly, the magnetic moment and corresponding $g$-factor of 1.952$\mypm$0.002 
is insensitive to  charge carrier type, concentration, and mobility. 
\end{abstract}

% \pacs{Valid PACS appear here}% PACS, the Physics and Astronomy
\maketitle

%\tableofcontents

Graphene is widely recognized as a promising material for spintronics applications due to its many favorable properties, such as tunable charge carrier concentration, high electronic mobility, and long spin diffusion lengths~\cite{vzutic2004spintronics,trauzettel2007spin,tombros2007electronic,candini2011graphene,han2014graphene,roche2014graphene}. Initial theoretical investigations into spin relaxation~\cite{fabian1999spin} for pristine graphene predicted microsecond spin lifetimes, but experimental results have so far 
determined lifetimes orders of magnitude shorter~\cite{tombros2007electronic,tombros2008anisotropic,han2009electrical,popinciuc2009electronic,jozsa2009linear,pi2010manipulation,han2011spin,jo2011spin}.

In order to address this discrepancy Mani~\textit{et al.}~\cite{mani2012observation} recently demonstrated the first electron spin resonance (ESR) in magneto-transport on single layer graphene. These types of ESR-measurements based on resistively detected spin-flips have already been proven successful early on in AlGaAs-heterostructures to determine the electronic $g$-factor in low-dimensional systems~\cite{stein1983electron}. This first investigation by Mani~{\it et al.} provided extremely valuable insights and allowed for the calculation of a magnetic moment and associated 
$g$-factor to be $ g_\parallel = 1.94 \pm 0.024 $. This is in contrast to the $g$-factor for the free electron, which underlines the charge carrier interaction in graphene. 

In this Letter we extend this first study of resistively detected ESR in graphene to include different charge carrier types, densities, and mobilities, by employing a field effect setup. In detail we study ESR with a Hall bar geometry realized with high-quality monolayer CVD graphene~\cite{asadi2015up,cao2009electronic,wang2011electrochemical,montanaro2016optoelectronic,wei2016mechanically,lyon2016}, hence enabling to test the charge carriers' spin coupling properties. 

The graphene is transferred to a \SI{300}{\nano\meter} layer of SiO$_{2}$ on top of degenerately-doped Si using a wet transfer process~\cite{li2009large,liang2011toward,lyon2016}
%add your own ref. 
in which a number of cleaning steps are performed in order to minimize organic and inorganic contaminants. The Ni/Au contacts on the graphene Hall bar was defined by photolithography and excess graphene is removed with an oxygen plasma. In order to minimize the amount of water and other adsorbates on the surface, the sample is annealed twice; once before mounting for 12 hours at \SI{350}{\celsius} in vacuum, and again for 48 hours at \SI{140}{\celsius} under vacuum after being mounted in the probe of the cryostat.

\begin{figure}
	\includegraphics[width=86mm]{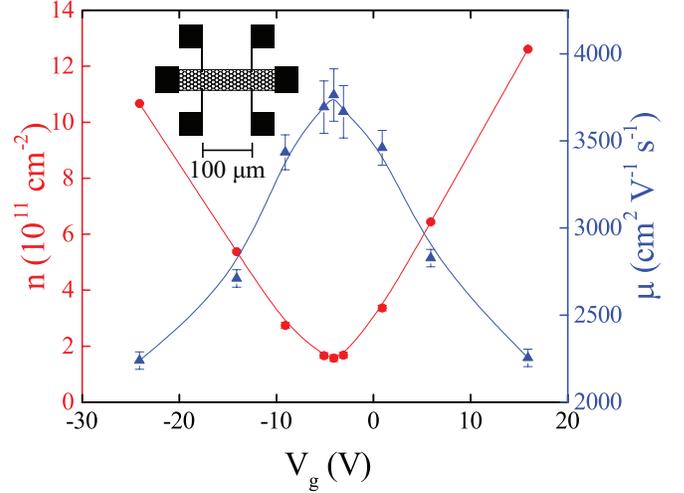}
	\caption{\label{fig:n_u} \footnotesize The charge carrier density (red) and mobility (blue) versus the gate voltage $V_{\rm g}$. The values are obtained by performing magneto-resistance measurements at the displayed values for $V_{\rm g}$. Points are connected by B-spline curves. Inset: A schematic of the Hall bar, which is \SI{200}{\micro\meter} long and \SI{22}{\micro\meter} wide.}
\end{figure}

Fig.~\ref{fig:n_u} shows the different charge carrier densities and mobilities measured at a temperature of \SI{4.2}{\kelvin}. Mobility at the charge-neutral point (CNP) is \SI{3760}{\centi\meter^{2}\volt^{-1}\second^{-1}}. The CNP appears at a gate voltage of $ V_{\rm g} = \SI{-4}{\volt} $.

%%%%%%%%%%%%%%%%%%%%%%%
%%%%%%%%%%%%%%%%%%%%%%%

\begin{figure}
	\includegraphics[width=86mm]{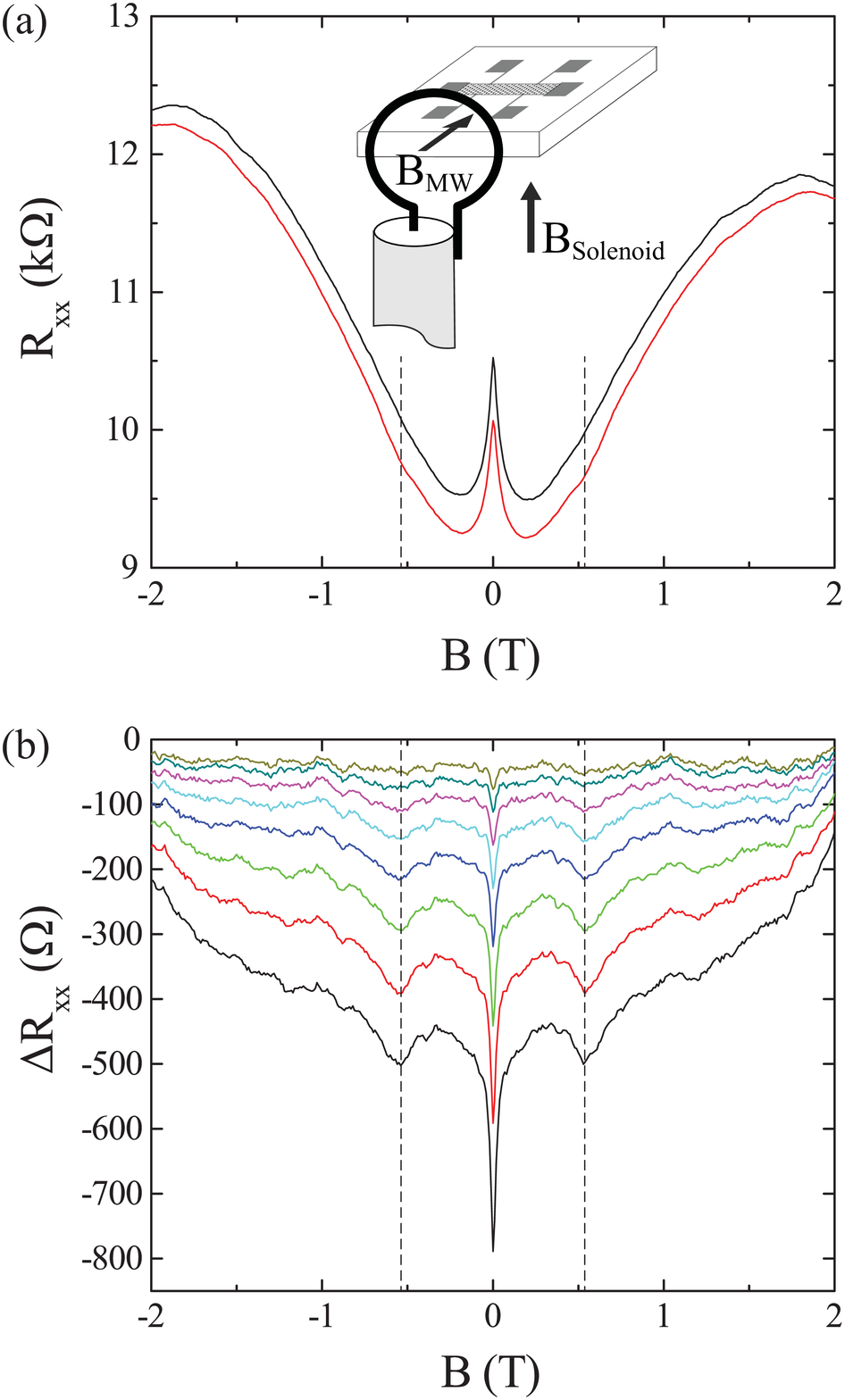}
	\caption[Graph displaying change in resistance with applied RF at different power levels]{\label{fig:diffpowercombined} \footnotesize (a) Measurements of longitudinal resistance both with (red, below) and without (black, above) applied RF at \SI{15}{\giga\hertz} and \SI{23}{dBm}, before subtraction. Inset: Diagram of experimental setup. The solenoid magnetic field is perpendicular to the plane of the Hall bar. Microwave radiation is applied in the plane of the sample and perpendicular to the applied current,
	coaxed by a cable connected to a loop antenna. (b) $ \Delta R_{\rm xx} $ for \SI{15}{\giga\hertz} RF signal at different applied RF power levels, with a back gate voltage of $ V_{\rm g} = \SI{-5.0}{\volt} $. Power levels are, from top to bottom, \SI{9} {dBm} to \SI{23}{dBm} in steps of 2~dBm. Note that the peaks remain stationary with increasing RF power which implies that there is no measurable temperature dependence for the $g$-factor. }
\end{figure}

Magneto-resistance measurements are made with two Stanford Research Systems SR830 Lock-in Amplifiers and microwave frequencies are generated with an Agilent E8257D signal generator. A schematic of the setup is shown in the inset of Fig. \ref{fig:diffpowercombined}. 
All measurements are performed at a temperature of \SI{4.2}{\kelvin} in a liquid helium cryostat with a solenoid magnet capable of generating a field of \SI{8}{\tesla}. The sample is mounted inside the solenoid such that the field is perpendicular to the plane of the Hall bar. A radio frequency (RF) signal is then applied to a loop antenna mounted to the end of a coaxial cable. A range of frequencies $ f $ between 15 and \SI{31}{\giga\hertz} which produce a standing wave were chosen for the measurements described below. This assures maximum RF power transmission to the graphene sample. Additionally, the charge carrier density of the graphene is varied by adjusting the back gate voltage. ESR measurements were performed for each frequency at nine different accessible charge carrier densities, around and including the CNP at $ V_{\rm g} = \SI{-4}{\volt} $.
Plots of the change in magneto-resistance induced by applying RF radiation are shown in Fig. \ref{fig:diffpowercombined}, with $\Delta R_{\rm xx} = R_{\rm xx}\textrm{(RF)} - R_{\rm xx}\textrm{(dark)}$.

\begin{figure}
	\includegraphics[width=86mm]{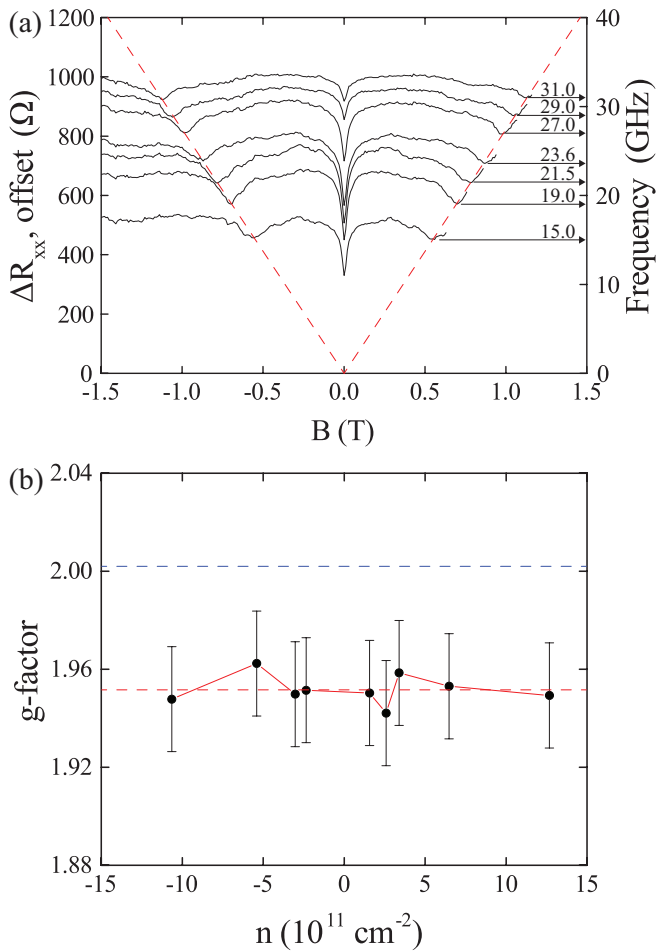}
	\caption{\label{fig:gfactor_combined} \footnotesize (a) $ \Delta R_{\rm xx} $ for different frequencies with the back gate set to $ V_{\rm g} = \SI{-4}{\volt}$. The measurements are offset to emphasize ESR peak shift due to $g$-factor. (b) Calculated $g$-factor at different charge carrier densities. Negative numbers represent holes. Red dashed line shows the mean value of $g$-factor measurements. Blue dashed line indicates $g$-factor of free electrons.}
\end{figure}

Following Mani~{\it et al.}~\cite{mani2012observation} we can attribute the resistance minima around $\mypm$\SI{0.5}{\tesla} to ESR. In this case, the applied radiation acts as a tipping field causing microwave-induced resonance with the charge carriers, which in turn leads to a decrease in resistance~\cite{bolotin2008temperature,heo2011nonmonotonic}. As displayed in Fig.~\ref{fig:diffpowercombined}, varying the applied RF-power does not result in differences in peak widths or positions. The weak localization peak around $B = \SI{0}{\tesla}$~\cite{beenakker1991quantum,suzuura2002crossover,morozov2006strong,khveshchenko2006electron,falko2007electron} naturally is smoothed out by RF irradiation. 
For each gate voltage, a linear relationship is found between the frequency of the applied microwave radiation and the magnetic field at which the resonance peak appears. This relationship is shown in Fig.~\ref{fig:gfactor_combined}(a), in which the sample is tuned to the charge neutral point, $ V_{\rm g} = \SI{-4}{\volt} $. The linear fit for this graph results in a slope of $ df/dB = (27.3 \pm \SI{0.3})\si{\giga\hertz~\tesla^{-1}} $. Employing the following relationship for the $g$-factor,
\begin{equation}
	 g = \frac{\mu}{\mu_{B}}, 
\end{equation}
with $\mu = h (df/dB)$ ($ \mu_{B} $ being the Bohr magneton) we find the absolute value of $\abs{g_\parallel} = 1.95 \pm 0.02 $. 

Measurements at different carrier densities give very similar $g$-factors, which are summarized in Fig. \ref{fig:gfactor_combined}(b). We find no dependence of the $g$-factor on charge carrier type, density, or mobility as shown in Fig. \ref{fig:gfactor_combined}.
The $g$-factors extracted for the nine different charge carrier concentrations have a mean value and standard deviation of $ \abs{g_\parallel} = 1.952 \pm 0.002 $ with the mean value within the uncertainty of every individual $g$-factor. Our results are compatible with the value of $ g_\parallel = 1.94 \pm 0.024 $ reported by Mani {\it et al.} assumed for holes in graphene placed on silicon carbide.
%
%%%%%%%%%%%%%%%%%%%%%%%
%%%%%%%%%%%%%%%%%%%%%%%
%

Earlier reports of ESR studies on graphite revealed a dependence of the $g$-factor on temperature as well as the external magnetic field orientation relative to the (out of plane) $c$-axis~\cite{matsubara1991electron,huber2004fluctuating}. These studies indicate that when $ B \perp c $-axis, $ g_{\perp} $ is essentially constant with respect to temperature and very close to that of the free electron, with $ g_{\perp} = 2.0023 $. In contrast, when $ B \parallel c $-axis, such as in our measurements there is an inverse relationship between $ g_{\parallel} $ and temperature, with $ g_\parallel $ ranging from around 2.05 to over 2.15 as the temperature is lowered from \SI{300}{\kelvin} to \SI{4.2}{\kelvin} due to changes in effective electron density. As can be seen from Fig.~\ref{fig:diffpowercombined}, peak locations in our measurements do not change with increasing RF power. Since increased microwave excitation causes a higher carrier temperature, we can deduce that there is no measurable effect on our measured $g$-factor due to temperature. 
The previously mentioned studies also predict that there is a $g$-factor shift, $ \Delta g $, in quasi-2D graphite that is either positive or negative as the dominant charge carrier changes between electrons and holes respectively, and $ \Delta g = 0 $ when the graphite is completely compensated.
This variance is not seen in our single layer graphene data, but one might expect charge carriers in graphene to have a different $g$-factor to that of graphite due to its properties such as Dirac-like linear dispersion relation and zero band gap, as well as possibly different QED corrections. 
Theoretical studies of charge carrier interaction at high magnetic fields predict an effective $g$-factor $g^{*}=2-4$ that oscillates depending on the filling factor $\nu$~\cite{volkov2012interaction}, and averages to a value of 2.3 in the presence of disorder.

Based on the resonance peak widths we now calculate the spin lifetime
\begin{equation}
	\tau_{\rm s} = \frac{\hbar}{2 \Delta E},
\end{equation}
where $ \Delta E = \mu \Delta B $ is the energy for the Zeeman splitting due to the applied magnetic field, and $ \Delta B $ is the resonance half-width of the ESR peaks. The resonance half-width is found to have no clear dependence on charge carrier density, applied RF power, or frequency outside of measurement error, and is generally found to be $ \approx (0.080 \mypm 0.015) $\si{\tesla}. This results in a spin relaxation time of approximately $ \tau_{\rm s} \cong (40 \mypm 6) $\si{\pico\second}. With this value, we then find the spin diffusion length at the CNP from 
\begin{equation}
	\lambda_{\rm s} = \sqrt{D \tau_{\rm s}} 
\end{equation}
to be $\lambda_{\rm s}\approx (630 \mypm 50)$~\si{\nano\meter}, where $D$ is the diffusion constant~\cite{tombros2007electronic}. These values are consistent with 
previous experiments~\cite{tombros2008anisotropic,han2009electrical,jozsa2009linear,han2011spin,jo2011spin,han2012spin}.

We now consider the possible origins of the deviation of our measured $g$-factor from the free-electron value, $\Delta g \simeq g - 2 \simeq -0.05$: 
\begin{itemize}
\item[(a)] Vacancy defects are likely present in our sample and those of Mani~\textit{et al.}, predicting $g$-factors in principle roughly within the correct range. 
The magnitude of magnetic moments due to vacancy defects varies depending on their type and other properties of the system~
\cite{ma2004magnetic,yazyev2007defect,santos2012magnetism,nanda2012electronic,wang2012magnetic}. 
Irradiation of graphene with Ar$^{+}$ ions~\cite{just2014preferential} can result in vacancy defects with $g=2.001-2.003$ while N$^{+}$ ion 
bombardment~\cite{ney2011irradiation} can implant ions that produce a $g$-factor of 2.
%{\large ****do you really want this last sentence?}
However, we observe no change in $g$-factor at all as a function of charge carrier density, as could be expected from charge carrier density dependent Kondo screening.

\item[(b)] Another possible source of $g$-factor deviation could be due to 
nanometer-sized ripples in graphene,  which appear even when deposited on a substrate such as SiO$_{2}$. 
Those have a maximum local strain of approximately 1\%~\cite{ishigami2007atomic}.
Also, first-principle calculations 
yield local magnetic moments of  $\approx1.5-2\mu_B$ for defects on graphene ripples under strain~\cite{santos2012magnetism}.
However, due to the essentially random nature of grain boundary orientation on CVD graphene it is unlikely that these ripples would be 
oriented in a way that would produce our $g$-factor signal.  

\item[(c)] Metal adsorbates could also play a role in the $g$-factor. 
%Some metal atoms could altere the $g$-factor and it is possible that some metal adsorbates 
These could be present in our samples due to the copper foil that the graphene was grown on. However, they should not be present in Mani's epitaxially grown samples. 
In addition, no theoretically predicted $g$-factor due to metal adatoms precisely matches our measured $g$-factor.

\item[(d)] Finally, substrate effects may play a role, but then we note that our results on Si/SiO$_{2}$ are compatible with measurements on 
SiC performed by Mani \textit{et al.} within experimental error, also ruling out the effect of the substrate.

\end{itemize}

We stress that the  sign of $\Delta g$ does not change when transitioning from electrons to holes, as it would be expected, and that our value 
is very close to the one obtained previously by Mani~{\it et al}. 
This finally suggests a more fundamental, intrinsic mechanism of spin relaxation, such as intrinsic spin-orbit coupling (SOC)~\cite{kane2005quantum}, which is known to strongly influence the $g$-factor. In a simple fashion, SOC in graphene has been modeled in the presence of orbitals with $d$-symmetry~\cite{gmitra2009band,konschuh2010tight}. These first-principles calculations have estimated the intrinsic SOC zero-field splitting to be as large as several \SI{10}{\micro\electronvolt} while the energy dispersion very close to the $K$-point remains linear. One can estimate the correction to the $g$-factor to be $\Delta g \simeq 4\lambda_I a^2/(\Phi_0\mu_B)$, with $\lambda_I$ being the intrinsic SOC due to $d$-orbitals, $a$ the lattice constant, $\Phi_0$ the magnetic flux quantum, and $\mu_B$ the Bohr magneton. The magnitude of the SOC may therefore account for the $g$-factor and its behavior that we have determined experimentally. As demonstrated by Kane \& Mele \cite{kane2005quantum}, 
renormalization group calculations yield an enhancement of the intrinsic SOC potential by an order of magnitude. Hence, our results support a similar intrinsic enhancement in graphene.

In summary, we have measured a resistively detected electron spin resonance in monolayer graphene. 
The associated absolute magnitude of the g-factor $ \abs{g_\parallel} = 1.952 \pm 0.002 $ is independent of charge carrier type or density to within our experimental accuracy. The deviation from the $g$-factor of free electrons and the insensitivity to external effects strongly suggest intrinsic spin-orbit coupling in graphene.

We would like to thank M. Prada, K. Vyborny, and L. Tiemann for fruitful discussions, as well as and M. I. Katsnelson and A. Liechtenstein for valuable insight. We acknowledge support by the Air Force Office for Scientific Research within the MURI'08 (FA9550-08-1-0337) 
and the Center for Ultrafast Imaging (CUI - Cluster of Excellence 1074) of the Deutsche Forschungsgemeinschaft (DFG).

\bibliography{esrtex}

\end{document}